\documentclass[12pt]{article}

\usepackage{latexsym}
\usepackage{epsfig,cite}
\usepackage{mathrsfs}
\usepackage{amssymb,amsmath,amscd}

\newcommand{\C}{\mathbb{C}}

\newcommand{\CP}{\mathbb{CP}}
\newcommand{\RP}{\mathbb{RP}}

\newcommand{\R}{\mathbb{R}}

\newcommand{\cN}{\mathcal{N}}

\newcommand{\p}{\partial}
\renewcommand{\Im}{\mathrm{Im}\,}

\newcommand{\rd}{\mathrm{d}}

\newcommand{\Li}{\mathrm{Li}}

\newcommand{\AdS}{\mathrm{AdS}}
\newcommand{\Pfaff}{\mathrm{Pfaff}}
\newcommand{\Vol}{\mathrm{Vol}}

\newcommand{\be}{\begin{equation}\label}
\newcommand{\ee}{\end{equation}}
\newcommand{\bea}{\begin{eqnarray}\label}
\newcommand{\eea}{\end{eqnarray}}

\begin{document}

\thispagestyle{empty}

\begin{center}
{\Large\bf
{\Large Amplitudes at Weak Coupling as Polytopes in AdS$_5$}
}\\
\vskip 1truecm
{\bf Lionel Mason$^{*}$ \& David Skinner$^{\dagger}$ \\
}

\vskip 0.4truecm
$^{*}${\it The Mathematical Institute,\\ 
24-29 St. Giles', Oxford, OX1 3LB,\\ 
United Kingdom}\\

\vskip .2truecm 
$^{\dagger}${\it Perimeter Institute for Theoretical Physics,\\ 
31 Caroline St., Waterloo, ON, N2L 2Y5,\\ Canada}\\

\end{center}

\vskip 1truecm 
\centerline{\bf Abstract} 

\medskip

We show that one-loop scalar box functions can be interpreted as
volumes of geodesic tetrahedra embedded in a copy of $\AdS_5$ that has
dual conformal space-time as boundary.  When the tetrahedron is
space-like, it lies in a totally geodesic hyperbolic three-space
inside $\AdS_5$, with its four vertices on the boundary. It is a
classical result that the volume of such a tetrahedron is given by the
Bloch-Wigner dilogarithm we show that this agrees with the standard
physics formul\ae\ for such box functions.  The combinations of box
functions that arise in the $n$-particle one-loop MHV amplitude in
$\cN=4$ super Yang-Mills correspond to the volume of a
three-dimensional polytope without boundary, all of whose vertices are
attached to a null polygon (which in other formulations is interpreted
as a Wilson loop) at infinity.

\newpage
\setcounter{page}{1}\setcounter{footnote}{0}

\section{Introduction}
\label{sec:intro}

A key property of planar scattering amplitudes in $\cN=4$ super
Yang-Mills is the existence of a `dual' copy of the conformal
group. While the usual conformal group acts on the compactification of
space-time, this dual conformal group acts on the compactified space
of region momenta.  

The dual conformal group first emerged in
studies~\cite{Usyukina:1992jd,Broadhurst:1993ib,Drummond:2006rz} of
various integrals that contribute to planar amplitudes, and was used
in~\cite{Bern:2007ct} to construct an ansatz for the four-particle 
amplitude at five loops. Its significance was greatly enhanced by the
work of Alday \& Maldacena~\cite{Alday:2007hr}, calculating the strong
coupling limit of planar amplitudes from the area of a minimal surface
(string worldsheet) embedded in a copy of five dimensional anti-de
Sitter space whose boundary is the compactified space of region
momenta. Dual conformal symmetry was then studied systematically at
weak coupling in~\cite{Drummond:2007au, Brandhuber:2008pf,
  Drummond:2008cr, Drummond:2008vq,Elvang:2009ya,Brandhuber:2009hk},
and its presence understood as a reflection of the integrability of
planar $\cN=4$ super Yang-Mills in the amplitude
sector~\cite{Beisert:2008iq,Berkovits:2008ic,Drummond:2009fd,Alday:2009ga,Alday:2009dv,Alday:2010vh}.
Infra-red divergences render this symmetry anomalous, but it is
conjectured to be an exact symmetry of the loop \emph{integrand}.

In~\cite{Hodges:2009hk}, Hodges introduced twistors for the dual
conformal group\footnote{Twistors can in general be defined to be the
  chiral (perhaps pure) spinors of any conformal group.}, naming them
\emph{momentum twistors} since they relate to the conformal geometry
of momentum space.  Hodges showed that NMHV tree amplitudes could be
represented as volumes of polytopes in momentum twistor space,
providing a geometric understanding of the cancellation of spurious
singularities present in individual BCFW terms. Subsequently, a
general representation in momentum twistor space was found,
conjecturally for all tree amplitudes and leading singularities
in~\cite{Mason:2009qx} using a contour integral over an auxiliary
Grassmannian. This Grassmannian integral is analogous and equivalent
to the Grassmannian representation of tree amplitudes and leading
singularities based on ordinary twistors for the space-time
superconformal group~\cite{ArkaniHamed:2009dn,ArkaniHamed:2009vw}.

At one-loop, Passarino-Veltman reduction can be used to decompose
four-dimensional scattering amplitudes into momentum space scalar
integrals usually represented as boxes, triangles and bubbles. For
$\cN=4$ super Yang-Mills only boxes appear~\cite{Bern:1994zx}.  Very
recently, these box functions have been studied using momentum
twistors~\cite{Hodges:2010}.  The purpose of this note is to show how
this analysis leads to a new geometric interpretation of box functions
as volumes of three-dimensional tetrahedra in the five dimensional
anti-de Sitter space introduced in~\cite{Alday:2007hr} that has
compactified region momentum space as boundary. When the region
momenta are all space-like separated, the tetrahedron lies in a
totally geodesic hyperbolic three-space (Euclidean $\AdS_3$) inside
$\AdS_5$.  It is a classical result that the volume of such an
ideal\footnote{A tetrahedron in hyperbolic space is \emph{ideal} if
  all four vertices lie at infinity.}  tetrahedron is given by the
Bloch-Wigner dilogarithm, and we show that this agrees with the
evaluation of the four-mass box function in the physics
literature~\cite{Usyukina:1992jd,Isaev:2003tk,Denner:1991qq}\footnote{While
the relation between box functions and volumes of geodesic tetrahedra
in hyperbolic space has been noted before (see {\it
  e.g.}~\cite{`tHooft:1978xw,Denner:1991qq,Davydychev:1997wa,Gorsky:2009nv}),
these authors took the hyperbolic space to be the mass-shell $p^2=m^2$
in momentum space.  Geometrically, the hyperboloid $p^2=m^2$ lies
entirely at infinity in the $\AdS_5$ considered in this paper, and its
definition breaks dual conformal invariance.}.

The scalar diagrams corresponding to box functions admit various
degenerations where one or more corners of the box become massless. In
these limits the box functions themselves diverge. This divergence can
be seen clearly in the geometry, arising from the divergent volume of
a tetrahedron when an entire edge goes to infinity (which occurs for
each massless corner). The regularisation procedure of Alday \& Henn
{\it et al.}~\cite{Alday:2009zm,Henn:2010bk} particularly naturally in
this context as it is equivalent to regularising the tetrahedron's
volume by bringing its vertices in from the infinity of $\AdS_5$ to
lie on a finite surface, a horosphere. This procedure has a
straightforward relationship with dimensional regularisation.

One loop amplitudes in $\cN=4$ SYM can be expressed as a \emph{sum}
over various different boxes~\cite{Bern:1994zx}, with coefficients
determined by their leading singularities~\cite{Britto:2004nc}. The
geometric point of view allows us to regard the sum of boxes as the
3-volume of the union of the constituent tetrahedra.  In the simplest
case of $n$-particle MHV amplitudes, we show that the corresponding
tetrahedra fit together to form a 3-dimensional polytope without
2-boundary, embedded in $\AdS_5$. This gives a coherent geometrical
realisation of the amplitude as the volume of a closed (piecewise
linear) 3-manifold. Recent work~\cite{Nima} suggests that this
correspondence between amplitudes and polytopes can be
extended to higher MHV degree.

\section{Box functions as tetrahedra in AdS$_5$}
\label{sec:tetra}

\begin{figure}
\begin{center}
	\includegraphics[height=45mm]{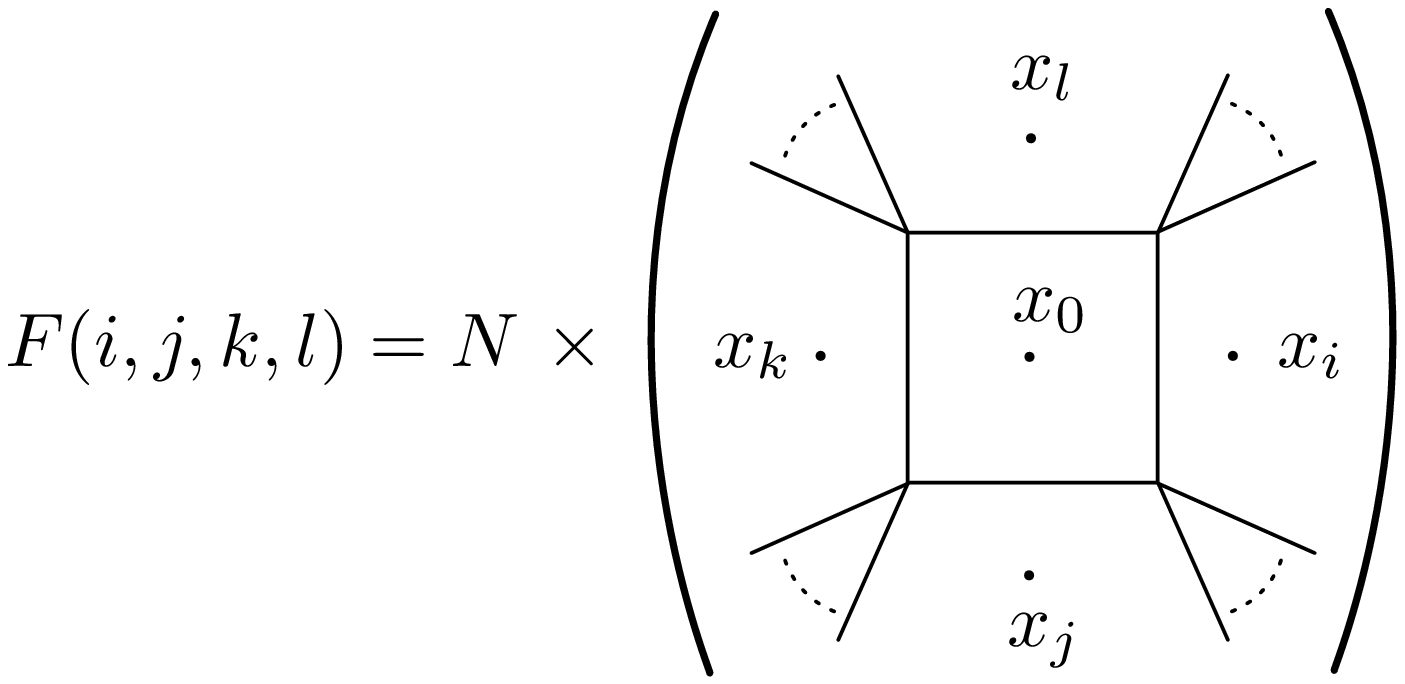}
\end{center}
\caption{{\it The 4-mass box function is defined by the integral~\eqref{4mbdef}. The region momenta are fixed up to overall translation by $x_i-x_{i+1} = p_i$, while the momentum running through a propagator is given by the difference of the region coordinates on either side of that propagator.}}
\label{fig:4mass}
\end{figure}

The external momenta $p_i$ of an $n$-particle colour-ordered Yang-Mills amplitude may be encoded in $n$ region momenta $x_i$, defined up to overall translation by 
\be{}
	x_i-x_{i+1}= p_i,\qquad \qquad x_{n+1}\equiv x_1
\ee
so that momentum conservation is automatic. In terms of region
momenta, the  4-mass box function is defined to be\footnote{The factor $-1/8\pi^2$ comes from the standard normalisation $-(4\pi)^{2-\epsilon}/(2\pi)^{4-2\epsilon}$ of the dimensionally regularised box function (in the limit $\epsilon\to0$ for the finite 4-mass box), together with the difference between our $N$ and the Gram determinant of {\it e.g.}~\cite{Bern:1994zx}.}
\be{4mbdef}
	F(i,j,k,l):=-\frac{N}{8\pi^2}\int_{\R^{3,1}}\frac{\rd^4x_0}
	{(x_{0i}^2+{\rm i}\varepsilon)(x_{0j}^2+{\rm i}\varepsilon)(x_{0k}^2+{\rm i}\varepsilon)(x_{0l}^2+{\rm i}\varepsilon)}\, ,
\ee
where $N$ is a normalisation factor (determined below) and $x_0$ is
the region momentum at the centre of the box (see
figure~\ref{fig:4mass}). Following standard practice, we will
implement the Feynman ${\rm i}\varepsilon$-prescription 
by analytically continuing the integrand
of~\eqref{4mbdef} to the space of complexified region momenta, and
rotating the contour of integration to the Euclidean real slice,
where we may set $\varepsilon=0$.  

It has long been known~\cite{Usyukina:1992jd,Broadhurst:1993ib} that when $\varepsilon=0$, the 4-mass box function is invariant under a `dual' conformal symmetry that acts on the region momenta $x$ in the same way as the usual conformal transformations act on space-time coordinates. This dual conformal symmetry becomes far more significant in the $\cN=4$ supersymmetric theory, as it is conjectured to extend (albeit anomalously) to the full planar amplitude~\cite{Drummond:2007au,Drummond:2008vq,Alday:2007hr} and may be viewed as the imprint of the integrability of the planar theory on scattering amplitudes~\cite{Drummond:2009fd,Alday:2010vh}. 

Dual conformal transformations do not act on the space $x$, because they can exchange a finite point with a point `at infinity'. To make invariance manifest, we will re-express the region momenta in terms of skew matrices $X^{\alpha\beta}$ ($\alpha,\beta=0,\ldots, 3$), taking
\be{Xcomps}
	x^{AA'}\rightarrow X^{\alpha\beta}
	=\begin{pmatrix} 
		-\frac{1}{2}\epsilon^{AB}x^2 & {\rm i}x^A_{\ B'}\\	
		-{\rm i}x_{A'}^{\ B} &  \epsilon_{A'B'}
	\end{pmatrix}\,.
\ee
Any matrix $X$ of this form satisfies
\be{KleinQ}
	X\cdot X:=\frac12 \epsilon_{\alpha\beta\gamma\delta}X^{\alpha\beta}X^{\gamma\delta}=0\,,
\ee
while for two such matrices $X_i$ and $X_j$ we have 
\be{dotprod}
	X_i\cdot X_j=-(x_i-x_j)^2\,.
\ee
(We work with a $(+---)$ signature space-time, so that $X_i\cdot X_j>0$ if $i$ and $j$ are space-like separated.) 

The special form of $X$ given in~\eqref{Xcomps} really represents a particular coordinate patch of (dual) conformally compactified space-time. The full, compacitified space may be thought of as the quadric $X\cdot X=0$ in $\RP^5$, on which $X^{\alpha\beta}\sim \lambda X^{\alpha\beta}$ are homogeneous coordinates. This quadric has signatures (2,4), (1,5) or (3,3) (and hence topology $S^1\times S^3$, $S^4$ or $(S^2\times S^2)/\mathbb{Z}_2$) in Lorentzian, Euclidean or ultrahyperbolic signature space-times, respectively. Dual conformal transformations act on $\RP^5$ via the vector field
\be{}
	J^{\alpha}_{\ \,\beta} =  X^{\alpha\gamma}\frac{\p}{\p
          X^{\gamma\beta}}\,, 
\ee
showing in particular that conformal transformations respect the linear structure of $\RP^5$ (whereas the linear structure of a Minkowski coordinate patch is not preserved).

The Feynman $i\varepsilon$ prescription can be satisfied by
analytically continuing the integrand to Euclidean signature and
setting $\varepsilon=0$.  We can therefore recast~\eqref{4mbdef} as
the contour integral 
\be{projbox} 
	F(i,j,k,l)= -\frac{N}{8\pi^2}\oint_{S^4}
	\frac{D^4 X}{X\!\cdot\! X_i\ X\!\cdot\!X_j\ X\!\cdot\!X_k\ X\!\cdot\!X_l}\,,
\ee 
where the holomorphic 4-form $D^4X$ is defined as 
\be{measuredef}
D^4X :=\frac1{\pi{\rm i}}\oint_{S^1} \frac {X\wedge\rd X\wedge\rd
  X\wedge\rd X\wedge\rd X\wedge\rd X}{X^2} 
\ee 
in terms of the canonical SL(6,$\C$)-invariant top form of weight 6 on
$\CP^5$. The $S^1$ contour in~\eqref{measuredef} is chosen to encircle the simple
pole at $X^2=0$, and so restricts the remaining integral to the
complex quadric $X\cdot X=0$.  The $S^4$ contour in~\eqref{projbox} is an
integration over (dual) conformally compactified Euclidean
space. Notice that we can replace the non-compact integral over
Euclidean space~\eqref{4mbdef} by an integral over $S^4$
because~\eqref{4mbdef} is regular as $x_0\to \infty$, reflecting the
UV finiteness of the original momentum space expression. The
normalization factor $N$ is fixed by the requirement 
\be{fix-norm}
	\frac{N}{(2\pi{\rm i})^4}\oint_{(S^1)^4} \frac{D^4X}{X\!\cdot\! X_i\ X\!\cdot\!X_j\ X\!\cdot\!X_k\ X\!\cdot\!X_l}=1\,, 
\ee 
where the contour is now taken to compute the residue at each simple pole in the denominator
(corresponding to cutting each of the four propagators in figure~\ref{fig:4mass}).  This integral is
straightforward\footnote{The integral may be performed by first choosing two vectors $E,F\in\C^6$ that are null, perpendicular to the span of $X_i$, and obey $E\cdot F=1$.  The affine coordinate chart of $\CP^5$ where $X\cdot F=1$ then has coordinates  $y_1=X\cdot X_i,\ y_2=X\cdot X_j,\ \ldots,\   y_5=X\cdot E$.  On this chart we can combine  equations~\eqref{measuredef} and~\eqref{fix-norm} to obtain the contour integral
$$
\begin{aligned}
	\frac{(2\pi{\rm i})^4}{N}&=\frac{1}{\pi {\rm i}}\frac{1}{|E\wedge F\wedge X_i\wedge X_j \wedge  X_k \wedge X_l|}
	\oint\frac{\rd y_1\wedge \rd y_2\wedge \rd y_3\wedge \rd y_4\wedge \rd y_5}
	{y_1y_2 y_3 y_4 (2y_5 E\cdot F+ Q(y,y))}\\
	&=\frac{1}{2\pi {\rm i}}\frac{1}{|X_i\wedge X_j \wedge  X_k \wedge X_l|}
	\oint\frac{\rd y_1}{y_1}\wedge \frac{\rd y_2}{y_2}\wedge\frac{\rd y_3}{y_3}
	\wedge\frac{\rd y_4}{y_4}\wedge\frac{\rd y_5}{y_5}
\end{aligned}
$$   
where $Q(y,y)$ is the part of the quadratic $X\cdot X$ involving only
$y_1,\ldots ,y_4$.  This now leads directly to \eqref{norm}.}  and
yields 
\be{norm} 
	N= \left|X_i\wedge X_j\wedge X_k \wedge X_l\right|
\ee 
where $|\cdot |$ denotes the norm induced on 4-forms from the flat
six-dimensional metric. Note that the sign of this normalisation
factor depends on the orientation of the contour used to
perform~\eqref{fix-norm}. This orientation is determined by an
ordering of the points $X_i,X_j,X_k$ and $X_l$.

\medskip

We can use the standard Feynman trick to reduce the denominator
of~\eqref{projbox} to a simple factor of degree 4.  Introducing four
real Feynman parameters $\alpha_a\in [0,1]=:I$, with $a=1,\ldots ,4$
one has 
\be{} 
	F(i,j,k,l)=-\frac{6N}{8\pi^2}\int_{I^4\times S^4}\hspace{-0.2cm}
	\rd^4\alpha\hspace{0.2cm}\delta\!\left(\sum_{a=1}^4\alpha_a-1\right)
	\frac{D^4 X}{(X\cdot X(\alpha))^4}\, , 
\ee 
where
\be{embed}
	X(\alpha):=\alpha_1 X_i+\alpha_2X_j+\alpha_3 X_k+\alpha_4X_l\, .  
\ee
In this $\RP^5$ framework, the individual propagators combine in a \emph{linear} fashion, compared to the quadratic expression one finds in momentum space.  The difference arises because in $\RP^5$, the propagator factors are linear in the homogeneous coordinates as a consequence of~\eqref{dotprod}.  In using the embedding in $\RP^5$, the tetrahedron of Feynman parameters, embedded via the $X(\alpha)$ of equation~\eqref{embed}, lies in the interior of $\AdS$ rather than being restricted to lie on the boundary $X\cdot X=0$ as in earlier work, for example~\cite{`tHooft:1978xw,Denner:1991qq,Davydychev:1997wa,Gorsky:2009nv}. As we will see, the linearity of $X(\alpha)$ in the $\alpha_a$ implies
that the embedding is totally geodesically for the $\AdS$ metric.

The $D^4X$ integral can be performed using momentum twistors. $X^2=0$ ensures that $X^{\alpha\beta}=A^{[\alpha}B^{\beta]}$ for some pair of linearly independent (momentum) twistors $A,B$.   We can thus replace the $X$ integral by an integral over $A$ and $B$ separately
\be{twistorint}
	\oint_{S^4}\frac{D^4X}{(X\cdot X(\alpha))^4}
	=\frac{1}{2\pi{\rm i}}\oint_{\CP^3}\frac{D^3A\, D^3B}{(A^\alpha B^\beta X_{\alpha\beta} (\alpha) )^4}
\ee
where the contour is over a copy of $\CP^3\subset\CP^3_A\times\CP^3_B$ defined by $A=\hat B$, where $\hat B$ denotes Euclidean complex conjugation
\be{}
	Z^\alpha= (Z^0,Z^1,Z^2,Z^3) \mapsto \hat Z^{\alpha} 
	= (-\overline{Z^1},\overline{Z^0},-\overline{Z^3},\overline{Z^2})\,.
\ee
Euclidean conjugation has no fixed points in twistor space, so on the contour $A=\hat B$, $A$ and $B$ never coincide and the integral on the right of~\eqref{twistorint} may be thought of as taken over the total space of the fibration
\begin{equation}
\minCDarrowwidth20pt
	\begin{CD}
	S^2 @> >>\CP^3\\
	@. @V VV\\ 
	@. S^4
	\end{CD}\quad ,
\end{equation}
thus encoding the original integral over conformally compactified Euclidean space (the factor of $1/2\pi{\rm i}$ in~\eqref{twistorint} compensates for the integration over the fibres;  see~\cite{Hodges:2010} for further discussion). This integral can be evaluated as 
\be{}
	\oint_{\CP^3}\frac{D^3A\,D^3B}{(A^\alpha B^\beta X_{\alpha\beta}(\alpha))^4}
	=\frac{1}{\det X(\alpha)}\oint_{\CP^3} \frac{ D^3 A\,D^3 C}{(A^\alpha C_\alpha)^4}
	=\frac{(2\pi{\rm i})^3}{6\det X(\alpha)}
\ee
where we have changed variables $B^\beta \to C_\alpha :=
X_{\alpha\beta}B^\beta$ at the expense of a Jacobian, and then used
the fact that the remaining integral is just ${\rm i}^3$ times the volume of $\CP^3$ in
the Fubini-Study metric. Since $X(\alpha)$ is skew symmetric 
\be{}
	\det X(\alpha)=(\Pfaff\, X(\alpha))^2 = \frac{1}{4}(X(\alpha)\cdot X(\alpha))^2
\ee
where we recall that the dot product is defined by the four index
$\varepsilon$-symbol that also defines the Pfaffian. Thus we obtain 
\be{Fproj2}
	F(i,j,k,l)=2\int_{I^{4}}\rd^4\alpha
        \;\delta\!\left(\sum_{a=1}^4\alpha_a-1\right)  
	\frac{\left|X_i\wedge X_j\wedge X_k\wedge
            X_l\right|}{(X(\alpha)\cdot X(\alpha))^2}\, . 
\ee
for the Euclidean 4-mass box integral. 

\medskip

\begin{figure}
\begin{center}
	\includegraphics[height=50mm]{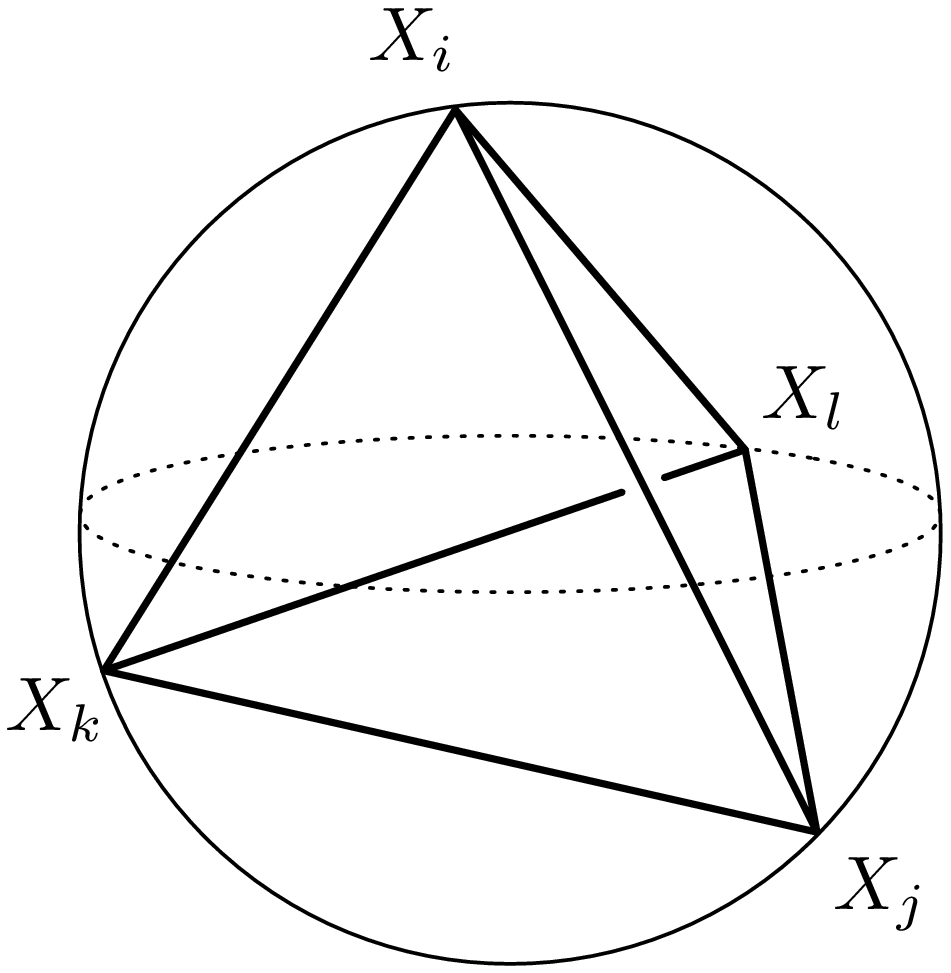}\hspace{3cm}
	\includegraphics[height=50mm]{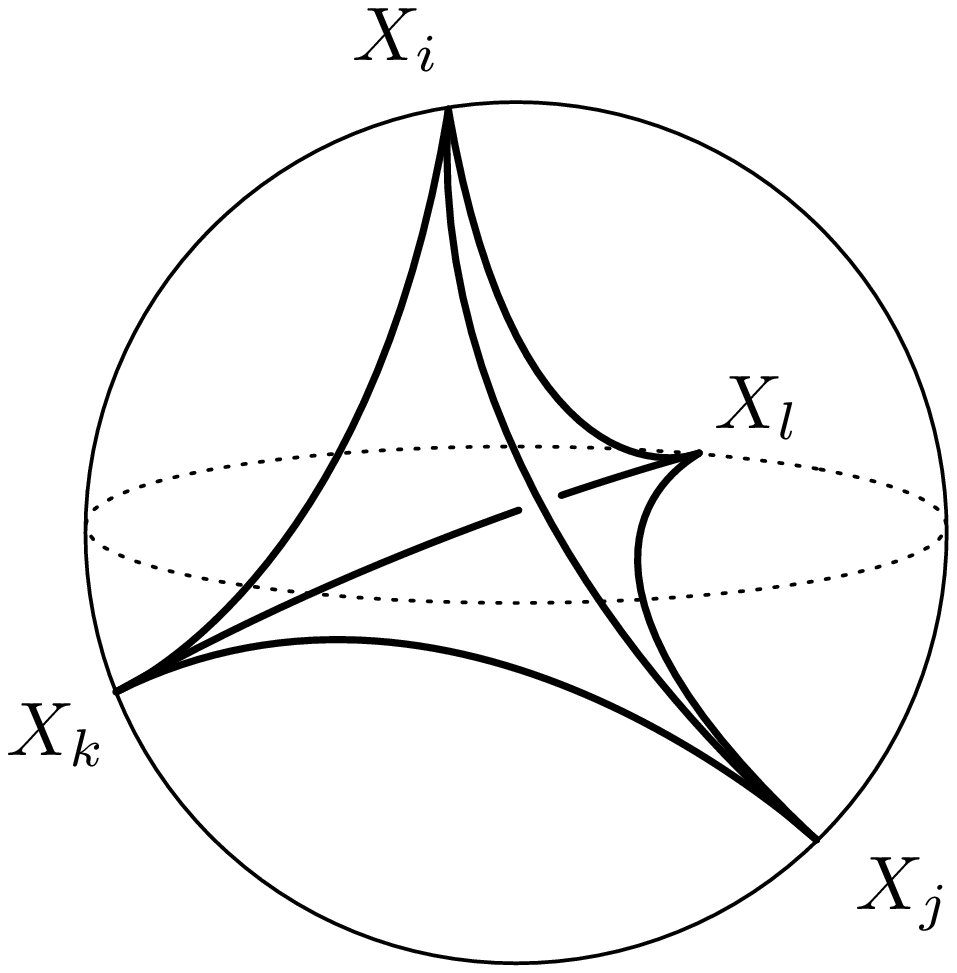}
\end{center}
\caption{{\it An ideal tetrahedron in AdS, shown in both the Klein--Beltrami (l) and Poincar{\' e} (r) models. All vertices lie on the conformal boundary at infinity, and each edge is an AdS geodesic.}}
\label{fig:tetra}
\end{figure}

We now turn to a geometric interpretation of this formula.  The expression 
\be{}
	X(\alpha)=\alpha_1X_i+\alpha_2X_j+\alpha_3X_k+\alpha_4X_l
\ee
defines a linear map from the space of Feynman parameters,
the unit simplex in $\R^3$, to
$\RP^5$. Provided the four vertices $X_i$, $X_j$, $X_k$ and $X_l$ are space-like separated (always true in Euclidean signature) we have $X(\alpha)\cdot X(\alpha)>0$, so we may take advantage of the
projective invariance of~\eqref{Fproj2} to introduce normalised coordinates 
\be{Ydef}
	Y(\alpha) = \frac{X(\alpha)}{\sqrt{X(\alpha)\cdot X(\alpha)}}
\ee
obeying $Y\cdot Y=1$. Thus $Y$ defines a map from the space of Feynman
parameters to Euclidean AdS$_5$, {\it i.e.}, the five dimensional
hyperbolic ball. Straight lines in $\RP^5$ are precisely the geodesics
in $\AdS_5$, so as the $\alpha_a$ vary over the 3-simplex
$\left\{\alpha_a\in\R^{\geq0}\ |\ \sum_{a=1}^4\alpha_a=1\right\}$,
$Y(\alpha)$ varies over a tetrahedron in $\AdS_5$ whose vertices lies
on the boundary at infinity, and whose edges and faces are totally
geodesic. Such a tetrahedron is called {\em ideal}, and is depicted in
figure~\ref{fig:tetra}.

We wish to show that $F(i,j,k,l)$ is simply (twice) the 3-volume of
this ideal tetrahedron.  Using the delta function to eliminate
$\alpha_4$, this volume can be written as the integral 
\be{vol1}
	\Vol(i,j,k,l)=\int_{\rm tetra} \rd^3\alpha \;
	\left|\frac{\p Y(\alpha)}{\p \alpha_1}\wedge\frac{\p Y(\alpha)}{\p \alpha_2}
	\wedge\frac{\p Y(\alpha)}{\p\alpha_3}\right| 
\ee
where $|\,\cdot\,|$ denotes the norm on 3-forms induced by the flat
metric in $\R^6\supset\AdS_5$.  Because $Y(\alpha)$ is normalised,
each derivative $\p Y(\alpha)/\p\alpha^a$ is orthogonal to $Y(\alpha)$
itself, so  
\be{}
	\left|\frac{\p Y(\alpha)}{\p \alpha_1}\wedge\frac{\p Y(\alpha)}{\p \alpha_2}
	\wedge\frac{\p Y(\alpha)}{\p \alpha_3}\right|
	\ =\ 
	\left|Y(\alpha)\wedge \frac{\p Y(\alpha)}{\p \alpha_1}\wedge
	\frac{\p Y(\alpha)}{\p \alpha_2}\wedge \frac{\p Y(\alpha)}{\p \alpha_3}\right| 
\ee
where now the norm is that on 4-forms.  The volume integral then reduces to
\be{Vol}
	\Vol(i,j,k,l)
	=\int_{\rm tetra}\rd^3\alpha\ \frac{\left|X_i\wedge X_j\wedge X_k\wedge X_l\right|}{(X(\alpha)\cdot X(\alpha))^2}\,,
\ee
which is exactly one half the expression for the 4-mass box function
obtained above. Since four generic\footnote{If the positions of the
  four vertices are not linearly independent, the normalization factor
  $\left|X_i\wedge X_j\wedge X_k\wedge X_l\right|$ vanishes and the
  volume is zero. This is obvious geometrically -- the tetrahedron is
  squashed flat.} points span a 3-space, any given tetrahedron in fact
lies inside some hyperbolic 3-space $\subset\AdS_5$. (When we
consider amplitudes in section~\ref{sec:amplitudes} it will be natural
to add various box functions together but it is then important to
remember that different tetrahedra lie in different three dimensional
subspaces.)

\medskip

Explicit formul\ae\ for this volume have been known since Lobachevskii~\cite{Milnor}.  It may be succinctly expressed as 
\be{Bloch-Wigner}
\begin{aligned}	
	B(z) &:= \Im \!\left\{\Li_2(z)\right\} + \arg (1-z) \log |z|\\
	&=\frac{1}{2}\left(\Li_2(z)-\Li_2(\bar z)\right) + \frac{1}{4}\left(\log(1-z)-\log(1-\bar z)\right)\log(z\bar z)\,,
\end{aligned}
\ee
where $B(z)$ is the `Bloch-Wigner dilogarithm' (and $\Li_2(z) =
-\int_0^z\log(1-t)\,\rd t/t$ is the usual `Spence' dilogarithm), whose
argument $z$ is interpreted as the cross-ratio of the location of the
four vertices on the boundary 2-sphere of the hyperbolic 3-space (see
{\it e.g.}~\cite{Zagier} for a review). The two-term dilogarithm
identities show that $B(z)$ depends on the ordering of the four points
in the cross-ratio only up to sign; equivalently, the sign of $B(z)$
depends on the orientation of the tetrahedron.  

We can make contact with the physics literature expression for the 4-mass box~\cite{Usyukina:1992jd,Isaev:2003tk} as follows.  We note first that the span of $X_i,X_j,X_k,X_l$  in $\RP^5$ is an $\RP^3$ and that generically,
$X\cdot X$ restricts to give a non-singular quadric (an $S^2$ in our Euclideanized setup) in this $\RP^3$.   We can choose homogeneous coordinates $y_i$, $i=1,\ldots,4$ on this $\RP^3$ so that 
\be{}
	\left.X\cdot X\right|_{\RP^3}= y_1y_2 -y_3^2-y_4^2\, .
\ee
In these coordinates, the natural  parametrization of points on the quadric by
a complex coordinate $\zeta$, defined up to a M{\"o}bius transformation,
is given by
\be{}
	(y_1,y_2,y_3,y_4)=(1,\zeta \bar \zeta, {\mathrm{Re}} \, \zeta, \Im \zeta)\, .
\ee
Since $X_i,X_j,X_k,X_l$ all lie on the quadric, they each correspond to a particular value of $\zeta$, and we can use the M{\"o}bius freedom to set 
\be{}
	(\zeta_i,\zeta_j,\zeta_k,\zeta_l)=(0,1,\infty,z)\, .
\ee
With these choices, we can now see that
\be{} 
	z\bar z=\frac{X_i\!\cdot\! X_l \ X_j\!\cdot\! X_k}{X_i\!\cdot\! X_j\ X_l\!\cdot\! X_k}\, ,
	\qquad \hbox{and}\qquad
	(1-z)(1-\bar z)=
	\frac {X_i\!\cdot\! X_k\ X_l\!\cdot\! X_j}{X_i\!\cdot\! X_j\ X_k\!\cdot\!  X_l} 
\ee

The Lorentzian formula may be obtained by analytic continuation of the
above Euclidean result, replacing $\bar z$ by $\tilde z$ and treating
$z$ and $\tilde z$ as independent complex variables. To do so one must
specifiy a branch of the dilogarithm, and the correct choice depends
on the channel of the scattering process (see {\it
  e.g.}~\cite{Denner:1991qq,Duplancic:2002dh} for details).

\section{Regularisation of lower-mass box functions}
\label{sec:lowermass}

\begin{figure}
\begin{center}
	\includegraphics[height=65mm]{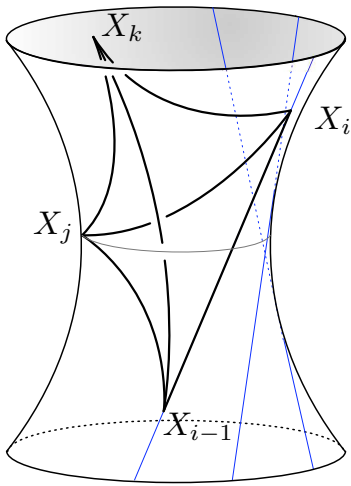}
\end{center}
\caption{{\it The 3-mass box integral has an entire edge along the boundary at infinity, so its volume diverges. This edge lies along a null geodesic (shown in blue) and the quadric $X\cdot X=0$ is ruled by these null lines.}}
\label{fig:3mass}
\end{figure}

Infrared divergences in $\cN=4$ super Yang-Mills amplitudes arise from
degenerations of the 4-mass box function when there is just a
single external massless particle attached to one or more corners of
the box (in Lorentzian signature). Since the vertices each lie on the
Klein quadric, the condition $0=p_{i-1}^2=(x_{i-1}-x_i)^2=X_i\cdot
X_{i-1}=0$ implies that 
\be{}
	(X_{i-1} + \alpha X_i)\cdot(X_{i-1} + \alpha X_i)=0
\ee
for any $\alpha$. Thus, an entire edge of the tetrahedra lies along
infinity, causing the volume to diverge (see figure~\ref{fig:3mass}).  
 
It is standard to regularise the divergence either by analytically
continuing to $4-2\epsilon$ dimensions, or by giving the external
states a non-zero mass. In the present context, Henn {\it et
  al.}~\cite{Alday:2009zm,Henn:2010bk} have shown that this latter mass
regularisation may be understood as moving onto the Higgs branch of
the theory. The resulting breaking of dual conformal invariance can be
understood geometrically in terms of moving the points on the quadric
$X\cdot X=0$ into the interior so that the geodesic tetrahedron that
they determine clearly has finite volume.  This can be done
systematically by projecting from a chosen point at infinity onto a
{\em horosphere} as follows.

\begin{figure}
\begin{center}
	\includegraphics[height=65mm]{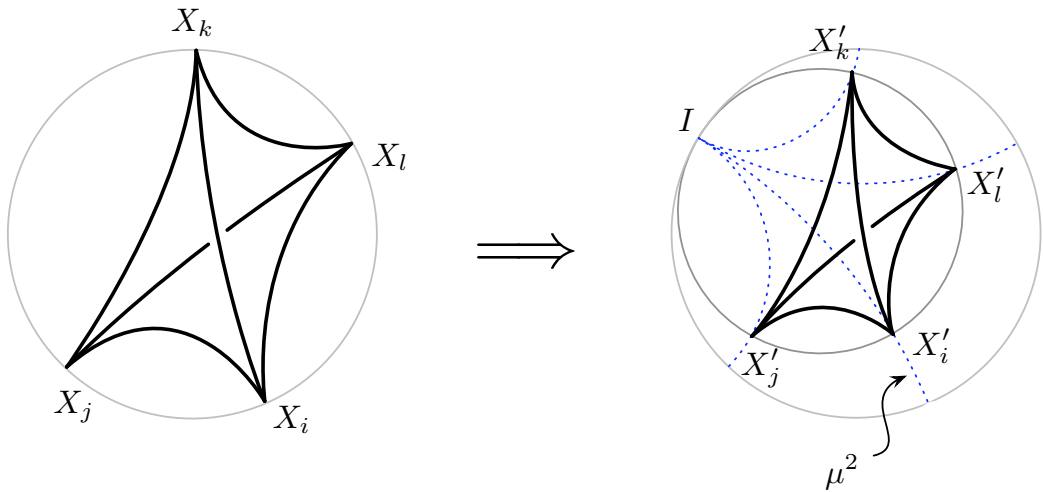}
\end{center}
\caption{{\it The tetrahedra may be regularised by
    bringing the vertices into the interior of AdS along the geodesic
    (shown in blue) connecting their original locations to a chosen
    point $I$. For simplicity, we can keep all the vertices on the
    same horosphere, corresponding to giving the external states equal
    masses.}} 
\label{fig:HorosphereTetra}
\end{figure}

Choose a point $I$ on the boundary of AdS that is space-like separated
from each of the original vertices. $I$ will be the point at infinity
in region momentum space and its choice breaks dual conformal
invariance. There is a unique AdS geodesic through $I$ and each
vertex and we can translate each vertex along the corresponding geodesic
by replacing 
\be{} 
	X_i \to X_i^\prime = X_i+\mu^2(I\cdot X_i) I 
\ee
where $\mu^2>0$ is a fixed parameter (see
figure~\ref{fig:HorosphereTetra}).  The points $X_i',\ldots X_l'$ then
lie on the horosphere $X\cdot X =2 \mu^2 (X\cdot I)^2$ in $\AdS_5$.

With this replacement, the regularised box functions are defined by the same integrals as before, so that ({\it e.g.}) the regularised 3-mass box integral is\footnote{The internal point $X$ still obeys $X\cdot X=0$.} 
\be{}
 F_{I,\mu^2}(i\!-\!1,i,j,k)=
	 2\int_{\overline M} D^4 X\ \frac{\left| X^\prime_{i-1}\wedge X^\prime_i\wedge X^\prime_j\wedge X^\prime_k\right|}
	 {X\!\cdot\! X^\prime_{i-1}\ X\!\cdot\!X^\prime_i\ X\!\cdot\!X^\prime_j\ X\!\cdot\!X^\prime_k}\ .
\ee
All the manipulations that relate this integral to the tetrahedron
3-volume~\eqref{Vol} (with vertices now at $X_i^\prime$ {\it etc.}) are
unaffected. Since the tetrahedron no longer reaches the boundary, its
volume is finite, but will depend on $I$ and $\mu^2$.  Once again, explicit
formul\ae\ for the volume of an arbitrary (non-ideal) tetrahedron and,
equivalently, for a scalar box function with arbitrary external masses
are known, both in the mathematical~\cite{ChoKim,MurakamiYano} and
physical~\cite{`tHooft:1978xw,Denner:1991qq,Davydychev:1997wa}
literature, and have recently been re-derived by
Hodges~\cite{Hodges:2010} from a point of view that is very close to
the present paper. Here we just give those cases relevant to the one
loop MHV amplitude.  Keeping only terms that do not vanish
as $\mu^2\to 0$, the regularised 2-mass easy box function
(figure~\ref{fig:MHV}) may be written as
\be{2me} 
\begin{aligned}
	& F_{I,\mu^2}(i\!-\!1,i,j\!-\!1,j) \ =\\
	&\ -\log\!\left(\frac{X_i\!\cdot\!X_j}{\mu^2\  X_i\!\cdot\!I\ X_j\!\cdot\!I}\right)
	\log\!\left(\frac{X_{i-1}\!\cdot\!X_{j-1}}{\mu^2\  X_{i-1}\!\cdot\!I\ X_{j-1}\!\cdot\!I}\right)
	+\frac{1}{2}\log^2\!\left(\frac{X_i\!\cdot\!X_{j-1}}{\mu^2\ X_i\!\cdot\!I\ X_{j-1}\!\cdot\!I}\right)\\
	&\  +\frac{1}{2}\log^2\!\left(\frac{X_j\!\cdot\!X_{i-1}}{\mu^2\  X_j\!\cdot\!I\ X_{i-1}\!\cdot\!I}\right)
	+\Li_2\!\left(1-\frac{X_i\!\cdot\!X_{j-1}\ X_{i-1}\!\cdot\! I}{X_{i-1}\!\cdot\!X_{j-1} \ X_i\!\cdot\!I}\right)
	+\Li_2\!\left(1-\frac{X_i\!\cdot\!X_{j-1}\ X_j\!\cdot\!I}{X_i\!\cdot\!X_j\ X_{j-1}\!\cdot\!I}\right)\\
	&\  +\Li_2\!\left(1-\frac{X_{i-1}\!\cdot\!X_j\ X_{j-1}\!\cdot\!I}{X_{i-1}\!\cdot\!X_{j-1}\ X_j\!\cdot\!I}\right)
	+\Li_2\!\left(1-\frac{X_j\!\cdot\!X_{i-1}\ X_i\!\cdot\!I}{X_i\!\cdot\!X_j\ X_{i-1}\!\cdot\!I}\right)
	 -\Li_2\!\left(1-\frac{X_i\!\cdot\!X_{j-1}\ X_j\!\cdot\!X_{i-1}}{X_i\!\cdot\!X_j\ X_{i-1}\!\cdot\!X_{j-1}}\right)
\end{aligned}
\ee
whereas the 1-mass box function is
\be{1m}
\begin{aligned}
	& F_{I,\mu^2}(i\!-\!3,i\!-\!2,i\!-\!1,i) \ =\\
	&\  -\log\!\left(\frac{X_{i-3}\!\cdot\!X_{i-1}}{\mu^2\ X_{i-3}\!\cdot\!I\ X_{i-1}\!\cdot\!I}\right)
	\log\!\left(\frac{X_{i-2}\!\cdot\!X_i}{\mu^2\ X_{i-2}\!\cdot\!I\ X_i\!\cdot\!I}\right)
	+ \frac{1}{2}\log^2\!\left(\frac{X_i\!\cdot\!X_{i-3}}{\mu^2\ X_i\!\cdot\!I\ X_{i-3}\!\cdot\!I}\right)\\
	&\ +\Li_2\!\left(1-\frac{X_i\!\cdot\!X_{i-3}\ X_{i-1}\!\cdot\!I}{X_{i-1}\!\cdot\!X_{i-3}\ X_i\!\cdot\!I}\right)
	+ \Li_2\!\left(1-\frac{X_i\!\cdot\!X_{i-3}\ X_{i-2}\!\cdot\!I}{X_{i-2}\!\cdot\!X_i\ X_{i-3}\!\cdot\!I}\right) + \frac{\pi^2}{6}
\end{aligned}
\ee 
and the zero-mass box function is 
\be{0m}
	F_{I,\mu^2}(1,2,3,4)=\log\!\left(\frac{X_1\!\cdot\!X_3}{\mu^2\ X_1\!\cdot\!I\ X_3\!\cdot\!I}\right)
	\log\!\left(\frac{X_2\!\cdot\!X_4}{\mu^2\ X_2\!\cdot\!I\ X_4\!\cdot\!I}\right) -\frac{\pi^2}{2}\ .  
\ee
We take the vertices of all our tetrahedra to 
lie on the same horosphere 
\be{}
	X\cdot X=2\mu^2(X\cdot I)^2
\ee
labelled by a single value of $\mu^2$.  These formul\ae\ may of course be simplified by normalising so that 
$X_i\cdot I=X_j\cdot I=\ldots=1$, but the expressions above serve to make the breaking of dual conformal invariance explicit.  Terms in~\eqref{2me}-\eqref{0m} that depend on $\mu^2$ are not invariant under a scale transformation of the boundary space-time, unless this is accompanied by a compensating rescaling of the regulator $\mu^2$. Terms that depend on cross-ratios involving the point $I$ are scale invariant, but not dual conformally invariant. 

The mass regularisation discussed here is related to the more commonly
used dimensional regularisation by a Mellin transform between the
parameter $\mu$ and the $\varepsilon$ of dimensional regularization in
dimension $4-2\varepsilon$; see appendix B of~\cite{Henn:2010bk} for
details.

\section{One loop MHV amplitudes}
\label{sec:amplitudes}

\begin{figure}
\begin{center}
	\includegraphics[height=40mm]{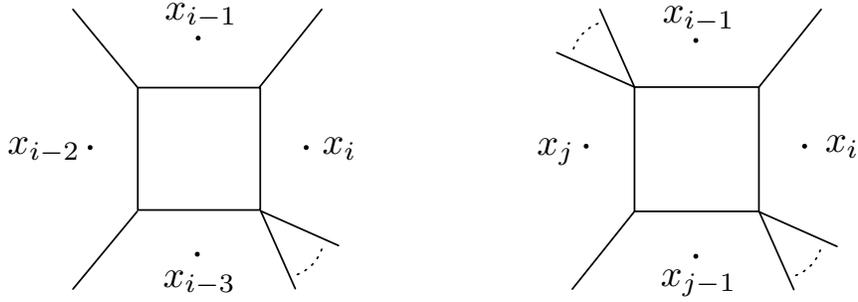}
\end{center}
\caption{{\it One loop MHV amplitudes involve only `1-mass' and `2-mass easy' box functions.}}
\label{fig:MHV}
\end{figure}

The one loop $n$-particle MHV amplitude in $\cN=4$ SYM may be
expressed~\cite{Bern:1994zx} as a sum of 1-mass and 2-mass easy box
functions (figure~\ref{fig:MHV}), each with coefficient one, times an
overall factor of the $n$-particle MHV tree amplitude. Since each box
function can be represented as a tetrahedron in $\AdS_5$, it is
natural to ask how these tetrahedra fit together to give a geometric
representation of the full one loop amplitude.

Individual 2-mass easy or 1-mass box functions correspond to
tetrahedra with vertices labelled by two consecutive pairs of integers 
\be{adjacentpairs}
	(i\!-\!1,i,j\!-\!1,j)\qquad\hbox{where}\qquad\{i,j\}\in\{1,\ldots,n\}
\ee
such that all four labels are distinct. There are thus $n$ 1-mass
boxes (where $j=i+2$ or $i=j+1$ working cyclically with $n+1=1$) and $n(n-5)/2$ 2-mass easy boxes.   
The orientation of the tetrahedra determined by the ordering of their
vertices in this list is thus unchanged under permutation of the {\it
  pairs} $(i\!-\!1,i)\leftrightarrow(j\!-\!1,j)$. This orientation
also induces an orientation on each face of a tetrahedron, for a
2-face is specified by a choice of three vertices, and we can declare
the normal to the face to be `inward' or `outward' pointing depending
on whether the omitted vertex lies in an odd or even position in the
ordering given by~\eqref{adjacentpairs}.

To construct the $n$-particle 1-loop MHV amplitude, we must combine
the volumes of these tetrahedra with coefficient one to obtain the
volume of the union of all the tetrahedra. Perhaps surprisingly, the
union is a 3-dimensional polytope without 2-boundary.  
The key point is that
each 2-face is shared by precisely two tetrahedra, each inducing
opposite orientations. In the generic case the 2-face $(i,j\!-\!1,j)$
is shared by tetrahedra $(i\!-\!1,i,j\!-\!1,j)$ and
$(i,i\!+\!1,j\!-\!1,j)$, each representing a 2-mass easy box. In the
first of these the omitted vertex $i 
\!-\!1$ is in an odd place in the labelling, whereas in the second,
the omitted vertex $i\!+\!1$ is in an even position, so the
orientations of the two tetrahedra are compatible. When
$(i,j\!-\!1,j)$ are all consecutive (so $j\!-\!2=i$), the second
tetrahedron is instead
$(i,j\!-\!1,j,j\!+\!1)=(i,i\!+\!1,i\!+\!2,i\!+\!3)$, representing a
1-mass box\footnote{Note that tetrahedron $(i,j\!-\!1,j,j\!+\!1)$ is
  not available when $j\neq i+2$ as it then represents a 2-mass hard
  box function, which does not contribute to the MHV amplitude.}. The
gluing is again compatible with the orientation of the individual
components.

The tetrahedra thus combine to form a closed\footnote{The four
  particle one-loop MHV amplitude, involving a single, zero mass box
  function, seems to be exceptional in this regard, although even this
  can be thought of as two copies of the same tetrahedron and so is
  still closed, albeit degenerate.} three dimensional polytope in
$\AdS_5$, whose total 3-volume is proportional to the 1-loop MHV
amplitude.  For example, the five particle amplitude consists of the
five possible 1-mass boxes and lies within an $\AdS_4\subset
\AdS_5$. The five corresponding tetrahedra join together to form the
boundary of a 4-simplex.  For six particles, one finds a polyhedron
with nine tetrahedral 3-faces that (for generic kinematics) cannot be
restricted to live in any lower dimensional subspace of $\AdS_5$.

These amplitude polytopes have $n$ null edges that lie entirely on 
boundary of $\AdS_5$ (forming the polygonal Wilson loop) and hence
their volume diverges. The regularisation discussed in
section~\ref{sec:lowermass} is compatible with the picture of gluing
the basic tetrahedra into a polyhedron provided we always use the same
horosphere (i.e., we use the same reference point $I$ and parameter
$\mu$ for each box).  Thus the regularized amplitude will be a
polytope with vertices brought in from infinity to lie on a
horosphere.  Performing the Mellin transform in the parameter $\mu$ as
in appendix B of~\cite{Henn:2010bk} will then give the full
dimensionally regularized amplitude.

\section{Discussion}

General, one-loop N$^k$MHV amplitudes in $\cN=4$ SYM are always
expressible as linear combinations of box functions, so a 3-volume
interpretation is still possible.   However, beyond MHV the situation
is complicated by the fact that different box functions have different
leading singularities as coefficients~\cite{Britto:2004nc}.  If one
factors the 1-loop MHV amplitude out of the complete superamplitude,
the ratio is conjectured to be dual conformally invariant and
finite~\cite{Drummond:2008vq}. Intriguing evidence has
emerged~\cite{Nima} that each finite combination of boxes that then
appears with a particular coefficient corresponds to the volume of a
3-polytope in $\AdS_5$. Furthermore, just as for the MHV amplitude
itself, these polytopes are believed to have no 2-boundary. 

\smallskip

Following the tree-level discussion in~\cite{Hodges:2009hk}, it should
be possible to understand the cancellation of spurious singularities
at one loop also.  The situation is more complicated here and there
are different types of singularities that arise, and that can be
spurious.  For example, the most dominant singularities of a box
function are are the infrared sungularities. These are associated to
edges which go to infinity when the pairs of vertices at each end
become null separated.  According to the standard infrared properties
of one loop amplitudes we have in dimensional regularization \be{}
M^{\mathrm{1-loop}}_n|_{\mathrm{IR}}=
-\frac1{\epsilon^2}M^{\mathrm{tree}}_n\sum_{i=1}^n
(X_i\cdot X_{i+2})^\epsilon \ee and so the physical singularities
correspond precisely to those edges connecting vertex $i$ to vertex
$i+2$, and those corresponding to the other edges must therefore be
spurious.

We expect that $\ell$-loop amplitudes can similarly be associated with
volumes of higher dimensional polytopes embedded in the $\ell$-fold
Cartesian product of $\AdS_5$.  The basic geometry associated to the
use of Feynman parameters leading to polydtopes in copies of $\RP^5$
would seem to extend strightforwardly to higher loops.  Again, the
scheme here is reminiscent of the description of NMHV tree amplitudes
as volumes of polytopes in momentum twistor
space~\cite{Hodges:2009hk}. Although it is clear that it should be
possible to describe N$^k$MHV amplitudes as volumes in the $k$-fold
product of momentum twistor space, it is simpler to use instead a dual
description in terms of residues in
Grassmannians~\cite{ArkaniHamed:2009dn,Mason:2009qx}. It therefore
seems plausible that the $\RP^5$ description of one loop boxes should
extend most naturally to a Grassmannian description of higher loop
integrals.

\vspace{1cm}

{\Large\bf\noindent Acknowledgements}

\medskip

\noindent It is a pleasure to thank Nima Arkani-Hamed, Freddy Cachazo, James Drummond, Johannes Henn and Andrew Hodges for many useful discussions, and the Faculty and Staff of IAS, Princeton for hospitality while this work was carried out. The work of DS is supported by the Perimeter Institute for Theoretical Physics. Research at the Perimeter Institute is supported by the Government of Canada through Industry Canada and by the Province of Ontario through the Ministry of Research $\&$ Innovation. The work of LM was financed in part by EPSRC grant number EP/F016654, see also {\\ \tt http://gow.epsrc.ac.uk/ViewGrant.aspx?GrantRef=EP/F016654/1}.

\bibliographystyle{JHEP}
\bibliography{DilogsTetraRefs}

\end{document}